\documentclass[11pt]{article}
\usepackage{graphics,tabularx,subfigure,color,calc,verbatim,epsfig,fullpage,authblk,appendix}
\usepackage[normalem]{ulem}
\usepackage{textcomp}
\definecolor{remarkcolour}{rgb}{0.8, 0.0, 0.0}
\usepackage{setspace} 
\usepackage[round,numbers,sort&compress]{natbib} 
\usepackage[font=small,labelfont=bf]{caption}
\usepackage{pdfpages}

\makeatletter
\renewcommand\@biblabel[1]{#1.}
\makeatother

\makeatletter
\def\@cite#1#2{$()#1\if@tempswa , #2\fi$}
\makeatother

\usepackage{lineno}

\title{Single-molecule stretching shows glycosylation sets tension in the hyaluronan-aggrecan bottlebrush}

\author[1]{S. N. Innes-Gold}
\author[1,$\dagger$]{J. P. Berezney} 
\author[1,2]{O. A. Saleh\thanks{Correspondence should be addressed to: saleh@ucsb.edu}}
\affil[1]{\footnotesize{University of California, Santa Barbara, Materials Department}}
\affil[2]{University of California, Santa Barbara, Biomolecular Science and Engineering Program}
\affil[$\dagger$]{Current affiliation: Brandeis University, Physics Department}

\begin{document}
\maketitle

\begin{abstract}
Large bottlebrush complexes formed from the polysaccharide hyaluronan (HA) and the proteoglycan aggrecan contribute to cartilage compression resistance and are necessary for healthy joint function. A variety of mechanical forces act on these complexes in the cartilage extracellular matrix, motivating the need for a quantitative description which links their structure and mechanical response. Studies using electron microscopy have imaged the HA-aggrecan brush but require adsorption to a surface, dramatically altering the complex from its native conformation. We use magnetic tweezers force spectroscopy to measure changes in extension and mechanical response of an HA chain as aggrecan monomers bind and form a bottlebrush. This technique directly measures changes undergone by a single complex with time and under varying solution conditions. Upon addition of aggrecan, we find a large swelling effect manifests when the HA chain is under very low external tension (i.e. stretching forces less than $\sim$1 pN). We use models of force-extension behavior to show that repulsion between the aggrecans induces an internal tension in the HA chain. Through reference to theories of bottlebrush polymer behavior, we demonstrate that the experimental values of internal tension are consistent with a polydisperse aggrecan population, likely caused by varying degrees of glycosylation. By enzymatically deglycosylating aggrecan, we show that aggrecan glycosylation is the structural feature which causes HA stiffening. We then construct a simple stochastic binding model to show that variable glycosylation leads to a wide distribution of internal tensions in HA, causing variations in the mechanics at much longer length-scales. Our results provide a mechanistic picture of how flexibility and size of HA and aggrecan lead to the brush architecture and mechanical properties of this important component of cartilage.
\end{abstract}

\noindent\fbox{%
    \parbox{\textwidth}{%
	SIGNIFICANCE \hspace{4pt} We use single-molecule stretching experiments to study a macromolecular complex of hyaluronan and aggrecan whose structure is crucial to maintaining mechanical strength of articular cartilage. We experimentally validate a model which quantitatively describes how the extension of an HA chain is affected by binding aggrecan side chains. This model yields information about the sensitivity of the complex size to different features of the bottlebrush architecture, and predicts when and how aggrecan damage leads to collapse of the complex. 
    }%
}

\section*{Introduction}

\begin{figure}[h!]
\centering
\includegraphics[width = 6.5in]{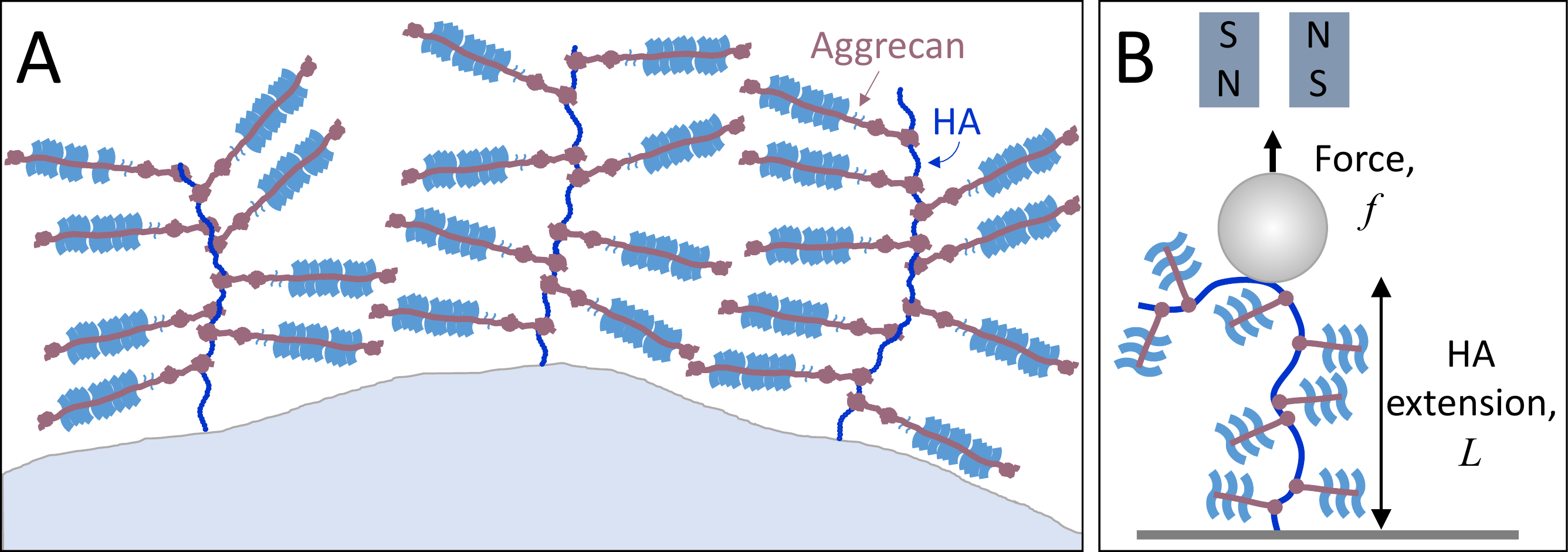}
\caption{A. Cartoon of the chondrocyte glycocalyx, wherein HA chains are tethered to the cell surface and decorated with the proteoglycan aggrecan, forming a microns-thick coating around the cell. B. Experimental set-up: a single HA chain, decorated by aggrecans, is tethered between a glass surface and a magnetic bead. A magnet applies stretching forces to the HA chain at the center of the bottlebrush.}
\label{fig:cartoon}
\end{figure}

Large bottlebrush complexes formed from the polysaccharide hyaluronan (HA) and the proteoglycan aggrecan constitute a major component of cartilage \cite{Kiani2002}. HA is a long, linear polyelectrolyte with a charge density of 1 \textit{e}/nm, a persistence length of around 5 nm, and a native size of $\sim$1-10 MDa, corresponding to contour lengths of $\sim$2-20 $\mu$m \cite{Cowman2005}. In cartilage, HA is secreted by chondrocytes and anchored to the cell surface by HA synthases or the receptor CD44 \cite{Bastow2008}, forming the sugar-rich ``glycocalyx'' coating around the cell. Each HA chain in the glycocalyx is complexed with many non-covalently bound aggrecans to form a large macromolecular bottlebrush that is known to be a major contributor to cartilage compression resistance \cite{Knudson2001}. In cartilage, the complex involves a third participant, link protein (LP), which binds to both HA and aggrecan to stabilize the interaction \cite{Knudson2001}. As illustrated in Fig. \ref{fig:cartoon}A, each aggrecan monomer is itself a bottlebrush, consisting of a $\sim$300 kDa protein core which is densely decorated by the glycosaminoglycan side chains keratan sulfate (KS) and the more abundant chondroitin sulfate (CS) \cite{Knudson2001}. The core protein includes three globular domains: two at the N-terminus (G1 and G2) and one at the C-terminus (G3) \cite{Kiani2002}. G1 is the HA-binding domain. In between G2 and G3 is an extended domain decorated by $\sim$100 CS chains \cite{Kiani2002}. The high charge and large size of aggrecan result primarily from this CS-rich region. This structure-defining domain changes with age, showing a reduction in the number and size of CS side chains \cite{Lee2013,Vasan1980,Watanabe1998,Hardingham1998,Dudhia2005,Ng2003}.

Previous studies have used electron microscopy and atomic force microscopy (AFM) to visualize the structure of individual aggrecans and aggrecan-HA complexes \cite{Morgelin1988,Morgelin1995,Neame1993,Yeh2004,Ng2003}. These studies have provided estimates of the length and diameter of aggrecan, as well as the density of bound aggrecans on an HA chain. However, imaging requires adsorption to a surface which significantly distorts the conformation of the complex. In order to learn about the solution structure of the complex, including the conformation adopted by the central HA chain, it is necessary to probe the structure of individual complexes by a method that does not confine the aggregates to 2D. To this end, others have used single-molecule force spectroscopy to measure properties of the complexes, such as inter-aggrecan interactions \cite{Harder2010,Han2007}. AFM and laser tweezers, which can access forces on the order of 10-1000 and 1-100 pN respectively \cite{Neuman2008,Saleh2015}, have also been used to test the strength of the bond connecting aggrecan and HA \cite{Fantner2006,Kienle2014,Liu2004again,Bano2018}. Liu \textit{et. al.} used laser tweezers to probe moderate-force tensile behavior ($\sim$10-40 pN) and compressibility of single HA-aggrecan complexes \cite{Liu2003,Liu2004}.

To our knowledge, no prior work has conducted low-force tensile testing on single bottlebrush complexes with sub-piconewton resolution, despite this likely being the biologically relevant force range: a study on osteocytes estimated that flow forces can lead to tension in the cell-attached components (e.g. HA) in the range of $\sim$1-10 pN \cite{Wang2007}. Here, we aim to quantify the structure of the complex based on its response to forces in this range. To do this, we use magnetic tweezers force spectroscopy. Prior work has shown that magnetic tweezers are particularly useful for quantifying structural changes that manifest at low forces, such as intra-chain electrostatic repulsion \cite{Jacobson2017}. Thermally dominated structure at a particular length-scale $x$ is probed by forces of the scale $k_BT/x$ \cite{Pincus1976,Saleh2015}. Magnetic tweezers typically apply forces on the order of 0.01-10 pN, and thus are sensitive to structure on length-scales between $\sim$1-100s of nm. Because of this long length-scale sensitivity, magnetic tweezers have succeeded in measuring the structural changes induced by side chains in a synthetic polymer bottlebrush \cite{Berezney2017}. 

By applying low-force magnetic tweezers stretching to study the formation and properties of the HA-aggrecan bottlebrush complex, we arrive at a novel physical description of this important component of cartilage. Prior work has shown that aggrecan swells HA chains in pericellular coats \cite{Chang2016} and pure HA brushes \cite{Attili2013}; we find this effect is also detectable at the single-chain level. Thus, we can observe in real time the assembly of the complexes by tracking the increase in chain extension. We find our data are well-described by a model in which the effect of all inter-aggrecan repulsions can be combined into a single internal tension, which acts in combination with the externally applied stretching force to extend the HA chain. We show that, on average, the internal tension generated in the central HA chain is about 0.4 pN, but with large variability between measurements. We use an analytical theory of bottlebrush physics to show that the spread in the data is consistent with variability in degree of glycosylation of the bound aggrecan.

\section*{Materials and Methods}

Hyaluronic acid (2.5 MDa), chemically functionalized to allow tethering to two surfaces, was purchased from Creative PEGworks. Each HA chain has a biotin group at the reducing end, and randomly situated thiol groups with a stoichiometry of one thiol per chain. Microfluidic sample chambers were constructed using maleimide-functionalized PEG-grafted glass coverslips purchased from Microsurfaces, Inc. HA chains were attached to the surface via reaction of the thiol groups with the maleimide, carried out in 50 mM sodium phosphate buffer (pH 7.2), 50 mM NaCl, 0.03\% Tween-20, and 100 mM TCEP. After attachment, excess polymer was removed by rinsing with 10 mM MOPS buffer (pH 7, with ionic strength of 4 mM). Streptavidin-coated magnetic beads, 1 $\mu$m in diameter (MyOne C1 paramagnetic beads, Invitrogen) were attached to the biotin-labeled chain ends. 

This experimental protocol leads to polydispersity in the tether lengths due to the random locations of the thiols, and necessitates normalization by contour length to compare measurements on different tethers. We expect an approximately uniform distribution of tether sizes up to $\sim$6 $\mu$m. However, due to the potential for bead-surface interactions at low forces, we do not collect data on tethers shorter than $\sim$0.6 $\mu$m. The stochastic thiol labeling strategy means that the magnetic bead is attached internally within the chain, and thus that there is a second aggrecan-decorated HA (not pinned to the surface, and so under no tension) in close proximity to the elastically-tested tether (see Fig. \ref{fig:cartoon}B). The excluded volume presented by this second chain is small compared to the bead and surface, and we do not expect it to have an effect at the forces probed here.

Aggrecan ($\geq$2.5 MDa, purified from bovine articular cartilage) was purchased from Sigma-Aldrich. It was centrifuged to remove large aggregates, and small contaminants were removed by filtration with a 100 kDa spin column. Recombinant aggrecan G1-IGD-G2 (with C-terminal 10-His tag) was purchased from Biolegend. Experiments were conducted in 100 mM NaCl, 1-10 mM MOPS (pH 7), and 0.03\% Tween-20. 

Experiments were conducted using a custom-built magnetic tweezers instrument, as described in detail elsewhere \cite{Ribeck2008}. Briefly, HA-tethered beads are imaged, and bead position is measured using an image analysis routine based on the bead's diffraction pattern. The output of that routine includes the tether length and lateral bead fluctuations; the latter are analyzed to estimate the applied tension \cite{Lansdorp2012}. To ensure each bead is tethered by a single HA chain, the tether length is measured as the bead is rotated. Multiple tethers become interwound during rotation, leading to a characteristic decrease in bead height \cite{Strick1998}.

Experiments on aggrecan-induced swelling used an aggrecan concentration of 2 mg/mL (approximately 0.8 $\mu$M). Above $\sim$1 mg/mL, the HA chain extension does not change appreciably with additional aggrecan (see Supporting Material, Fig. S1). A concentration of 2 mg/mL is well into the concentration insensitive regime, and is close to the overlap concentration at which the distance between aggrecans in solution approaches the aggrecan radius of gyration \cite{Papagiannopoulos2006}. This relatively high solution concentration is still well below the density of aggrecan bound in cartilage, thought to be on the order of 10s of mg/mL \cite{Han2008}. For experiments using the G1-IGD-G2 fragment, the protein was dissolved at $\sim$0.8 $\mu$M. 

To obtain deglycosylated aggrecan core protein, aggrecan (Sigma-Aldrich) was incubated overnight with chondroitinase ABC (Recombinant \textit{P. vulgaris} Chondroitinase ABC with N-terminal Met and 6-His tag, purchased from R\&D Systems). The reaction took place at 37$^o$C in 50 mM Trizma HCl (pH 8), 60 mM sodium acetate, 0.02\% BSA. Chondroitinase ABC digests HA in addition to chondroitin sulfate, so it was necessary to separate it from the deglycosylated aggrecan before experiments could proceed. Separation was accomplished using Dynabeads His-Tag Isolation and Pulldown, which were then removed magnetically. The presence of aggrecan core protein ($\sim$300 kDa) at the expected concentration was confirmed with SDS-PAGE (Fig. S2) and Dynamic Light Scattering on denatured aggrecan (Fig. S3) using a Malvern Zetasizer Nano ZS instrument. HA force-extension curves in the presence of aggrecan core protein were obtained in 100 mM NaCl, 10 mM MOPS (pH 7), 0.03\% Tween-20, and 1 mg/mL free chondroitin sulfate (chondroitin sulfate sodium salt from shark cartilage, purchased from Sigma-Aldrich). The free chondroitin sulfate was included in order to protect the HA tethers from digestion by any residual enzyme, and was shown to have no effect on the HA mechanical response (Fig. S4).

\section*{Results}
\begin{figure}[h!]
\centering
\includegraphics[width = 6.5in]{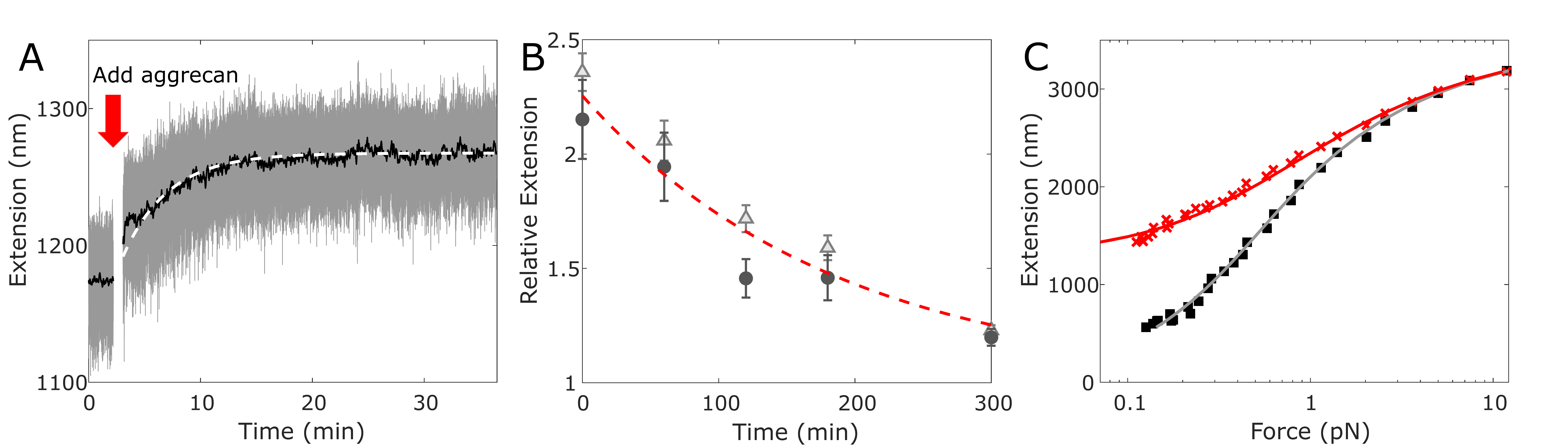}
\caption{A. HA extension while aggrecan binds (gray: all points; black: moving average over 4 seconds) under a constant applied force of 1.8 pN. The data are fit by an exponential with a characteristic timescale of 244 $\pm$ 2 seconds (white dashed line). B. HA extension during aggrecan unbinding. After removing free aggrecan from the bulk solution, extension of two aggrecan-decorated tethers at 0.2 pN (first tether: light gray triangles, second tether: dark gray circles), relative to aggrecan-free extension of the same tether at the same force, slowly decreases towards 1. Error bars show uncertainty arising from interpolation of force-extension curves at 0.2 pN. The data are fit by an exponential with a characteristic timescale of 190 $\pm$ 50 minutes (red dashed line).  C. Example force-extension data on a single HA tether before (black squares) and after (red x's) addition of 2 mg/mL aggrecan and equilibration of at least 15 minutes. Both force-extension curves are globally fit by Eq. 1, with common contour length $L_C$ and persistence length $l_p$. Best-fit parameters are $L_C = 3573 \pm 14$ nm, $l_p = 7.4 \pm 0.4$ nm, and  $f_{int} = 0.40 \pm 0.03$ pN, where $f_{int}$ is the internal tension generated by aggrecan ($f_{int}$ is set to zero for the data measured without aggrecan). Errors reflect 95\% confidence intervals for the fit parameters.}
\label{fig:data}
\end{figure}

With magnetic tweezers, we measure the extension of tethered HA chains as aggrecan is added to solution. Our data show that aggrecan expands the HA chain upon binding (Fig. \ref{fig:data}). When the chain is minimally stretched, we can track the progress of aggrecan binding by monitoring the chain end-to-end extension. As can be seen in Fig. \ref{fig:data}A, the length increases upon addition of 2 mg/mL aggrecan, and levels off after about 15 minutes, indicating the system has equilibrated. We can subsequently rinse free aggrecan out of the bulk solution and observe the HA length change as aggrecan unbinds (Fig. \ref{fig:data}B). This process takes several hours. Comparing timescales obtained by fitting exponential models to the data, we estimate that the timescale of aggrecan binding is at least 50 times faster than aggrecan unbinding.
While aggrecan interactions may lead to anti-cooperative behavior and non-exponential kinetics, the data (within the experimental resolution) appear exponential. Thus, to good approximation, a simple independent binding picture can be used, which here indicates that, since the off-rate is 50x slower than the on-rate, the maximal aggrecan density on HA has been achieved. We note that in cartilage, link protein would further stabilize the bound state, likely causing higher aggrecan density and even slower unbinding.

\begin{figure}[h!]
\centering
\includegraphics[width = 3.25in]{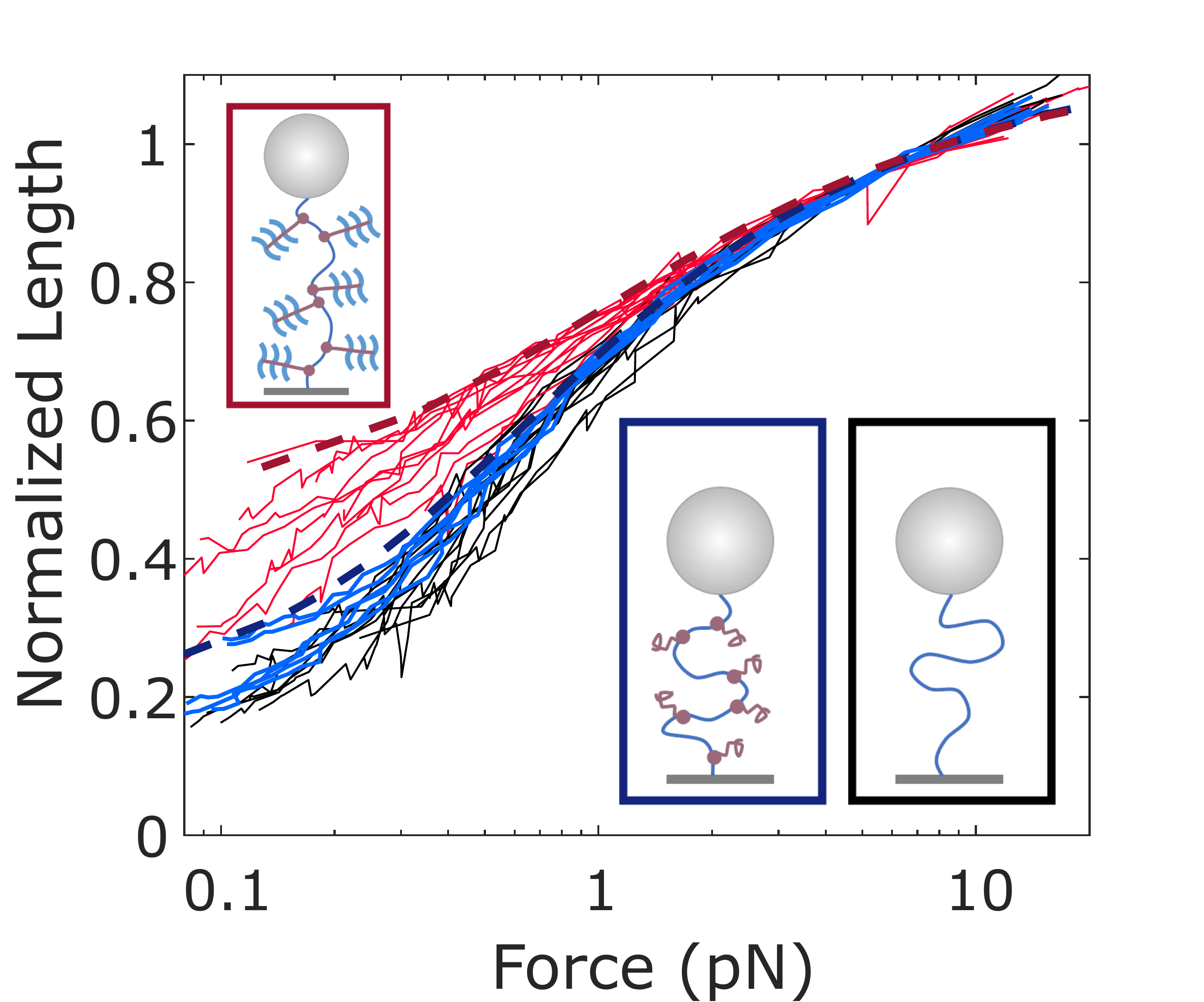}
\caption{Thin red lines show force-extension curves on 10 HA tethers with 2 mg/mL aggrecan. Extensions are normalized by multiplicative adjustment, with factors determined by comparing each curve to the median of the ten, and minimizing the sum of squared residuals in length for forces between 3 and 10 pN. Black lines show force-extension curves on the same 10 tethers in aggrecan-free conditions. Blue lines show force-extension curves of 6 HA tethers in the presence of deglycosylated aggrecan core protein. Thick dashed lines are predictions from the WLC-internal tension model for fully glycosylated aggrecan (dark red) and aggrecan core protein (dark blue). Cartoons (inset) illustrate the conditions for each set of force-extension curves (full aggrecan: red box, upper left; chABC-digested aggrecan: blue box, bottom middle; bare HA: black box, bottom right).}
\label{fig:tension}
\end{figure}

Relative to the mechanical behavior measured before the addition of aggrecan, the ligand-binding dramatically alters the low-force response (see example data in Fig. \ref{fig:data}C). Repeated experiments on 10 HA tethers, with and without aggrecan, all show similar elasticity changes (Fig. \ref{fig:tension}). With the addition of aggrecan (2 mg/mL), all tethers undergo swelling at low forces; this is reflected by the shallower slopes of the force-extension curves. The slope change indicates that aggrecan binding stiffens the chain, causing it to become less responsive to the externally-applied force. The observed low-force elasticity change is consistent with long length-scale structural effects: swelling and stiffening of the central HA chain due to repulsion between bound aggrecans, as expected from the charged brush structure of the aggrecan monomers \cite{Nap2008}.

We quantify the low-force swelling using an internal tension model \cite{Jacobson2017}, in which the long length-scale stiffening (caused by the bottlebrush architecture) is described by a mean-field internal tension $f_{int}$, which acts in addition to the applied force $f$ to straighten the polymer. This first-order correction captures the deviation from wormlike chain elasticity, which is most prominent at low forces where the chain physics is dominated by long-range side-chain interactions. The internal tension term is incorporated into the Marko-Siggia wormlike chain (WLC) model \cite{Marko1995}, giving:

\begin{equation}
f =  \frac{k_{B}T}{l_p} \left( \frac {1}{4} \left (1-\frac {L}{L_{C}} \right)^{-2}-\frac {1}{4}+\frac {L}{L_{C}} \right) - f_{int}
\label{eq:WLC}
\end{equation}

where $l_p$ is the intrinsic HA persistence length, $L$ is the chain extension, and $L_C$ is the contour length. Eq. \ref{eq:WLC} is used to fit force-extension data, with $L_C$ and $l_p$ fit globally for pairs of curves on the same molecule. $f_{int}$ is fixed to zero for the bare HA force-extension data (example: Fig. \ref{fig:data}C). We repeated such analyses on each of the ten tethers, measuring force-extension curves before and after the addition of 2 mg/mL aggrecan, and extracting $f_{int}$ as a fit parameter. The average was 0.35 $\pm$ 0.21 pN (standard deviation). The spread in $f_{int}$ reflects variability in the response of the aggrecan-decorated chains, demonstrated by the scatter of the red traces in Fig. \ref{fig:tension}. We hypothesize that this variability arises due to differences in glycosylation of the aggrecan samples, which originated in biological tissue \cite{Watanabe1998,Matthews2002,Lee2010}. 

For comparison, we digested the CS side chains of the aggrecan using chondroitinase ABC, and found that the resulting deglycosylated protein induces almost no swelling in HA (blue lines in Fig. \ref{fig:tension}). While the less abundant KS side chains are thought not to be affected by this enzyme, others have shown that in contrast to CS, removal of the KS side chains has little to no effect on the aggrecan monomer structure \cite{Lee2013}.

\section*{Discussion}

The HA-aggrecan force-extension data highlight that properties of the side chains play a crucial role in defining the bottlebrush architecture. At low forces ($\sim$0.1 pN), the end-to-end extension of HA molecules often increases more than twofold upon introduction of aggrecan. However, when HA is decorated instead by deglycosylated aggrecan, this bottlebrush-induced stiffening is dramatically reduced, demonstrating the importance of the chemical details of aggrecan. Even without artificial reduction of aggrecan glycosylation, the experiments display a wide range of responses. These observations can be explained using established polymer bottlebrush theory. We show the full range of stiffening observed is consistent with predictions from scaling theory combined with previous experimental measurements of HA and aggrecan structure. We further find that the variance in the glycosylated aggrecan data is consistent with an aggrecan population of varying degrees of glycosylation. Such results demonstrate the interplay between the branched architecture of the complex and the conformational structure of its constituent chains.

Following the logic of Berezney \textit{et. al.} \cite{Berezney2017}, we use a model developed by Panyukov \textit{et. al.} \cite{Panyukov2009} to predict, independently of our own measurements, the range of internal tensions in the HA backbone of bottlebrush complexes extended by glycosylated and deglycosylated aggrecan. The model treats the side chains as random walk polymers whose size, flexibility, and binding density control the internal tension generated in the central chain:

\begin{equation}
f_{int} = \alpha k_BT d^{\mu} b_a^{-(\mu+1)} n^{\nu} |\tau|^{t}
\label{eq:brush}
\end{equation}

where $\alpha$ is a prefactor that we take to be 1, $\mu$, $\nu$, and $t$ are scaling exponents (described below), $b_a$ is the Kuhn length (statistical segment length) of aggrecan, $n$ is the number of statistical segments within each aggrecan side chain, $\tau$ describes solvent quality and is 1 in good solvent, and $d$ is the distance between bound aggrecans. Estimates of $d$ range from 12-40 nm \cite{Neame1993, Yeh2004}; we choose the intermediate value of this range: 26 nm. If link protein were present, a smaller value ($\sim$10 nm) would be appropriate as link protein is known to increase aggrecan density \cite{Neame1993}. Based on our experiments showing unbinding is very slow, we assume the number and spacing of bound aggrecans are constant over experimental timescales (measurement of each force-extension curve takes 20 minutes). We assume that deglycosylation will be reflected as a change in aggrecan's Kuhn length $b_a$. The model (Eq. \ref{eq:brush}) considers the athermal limit and takes the exluded volume to be $b_a^3$. For bottlebrushes, Kuhn length typically scales with the diameter of the complex \cite{Birshtein1987}. Thus for fully glycosylated aggrecan, we will approximate $b_a \approx 50$ nm, roughly the diameter of the cylindrical proteoglycan \cite{Yeh2004}. In the opposite limit of complete deglycosylation, the side chains are aggrecan core protein alone. Based on the disordered nature of much of the core protein \cite{Jowitt2010}, we estimate $b_a$ as 1.6 nm, twice the persistence length of an unfolded polypeptide \cite{Rosales2012}. This value represents a physical lower limit; in reality it is likely that some residual structure leads to a slightly larger Kuhn length \cite{Jowitt2010}, but the model is not very sensitive to this assumption (Fig. S6). The number of segments $n$ is obtained by dividing the aggrecan contour length, estimated as 375 nm following AFM studies, \cite{Yeh2004,Ng2003} by the Kuhn length. SDS-PAGE of the deglycosylated protein (Fig. S2) shows a number of bands near the expected weight, suggesting there may be a spread of aggrecan sizes present; we do not incorporate this into our model but recognize that it may be an additional source of variability.

We assume the central HA chain is always strongly stretched due to its stiffness ($b_{HA} \approx$ 10 nm), side chains, and the external force, thus, the random coil structure of the side chains will be the dominant contributor to the HA internal tension. It is expected that deglycosylation will significantly perturb aggrecan structure; this will change not only its Kuhn length as discussed above, but also the scaling laws describing its solution structure. Fully-glycosylated aggrecan is strongly extended due to repulsion between its own CS branches. Following Panyukov \textit{et. al.} \cite{Panyukov2009}, the case of swollen aggrecan side chains dictates $\mu = -13/8$, $\nu = 3/8$, and $t = 1/8$ in Eq. \ref{eq:brush}. 

The deglycosylated core protein structure will also likely be dominated by the large CS-attachment region, now devoid of CS. Prior evidence has shown that the CS-attachment domain of aggrecan is structurally disordered \cite{Jowitt2010}. The solution structure of intrinsically disordered proteins reflects a trade off between conformational flexibility and transient short length-scale interactions, and is frequently modeled by ideal scaling \cite{Marsh2010}. Thus, we assume that without CS to provide stretching, the core protein will behave ideally, dictating scaling exponents $\mu = -5/3$, $\nu = 1/3$, and $t = 0$. We note that the model of Panyukov \textit{et. al.} is formally correct for a bottlebrush whose backbone and side chains have the same chemistry, which is not the case here. However, as discussed above, HA elasticity is not expected to have significant effects because the chain is always in a regime of strong stretching. Further, if we instead assume the core protein is swollen, the predicted tension does not change drastically (Fig. S6).
  
This model (Eq. \ref{eq:brush} with parameters listed above) allows us to estimate the range of $f_{int}$ values expected for aggrecan in different conditions: 0.15 pN in the limit of complete deglycosylation, and 0.51 pN for a monodisperse sample of undamaged aggrecan. Combining these values with the WLC model (Eq. \ref{eq:WLC}), we obtain expected force-extension curves for HA decorated by whole aggrecan (dark red dashed curve in Fig. \ref{fig:tension}) and HA decorated by deglycosylated aggrecan core protein (dark blue dashed curve in Fig. \ref{fig:tension}). Our experimental force-extension curves for aggrecan-decorated HA (Fig. \ref{fig:tension}) mostly fall between these limits, where the experimental curves showing the largest swelling are consistent with the undamaged aggrecan prediction of $f_{int} \approx$ 0.5 pN. The force-extension curves on HA with deglycosylated aggrecan core protein (blue traces in Fig. \ref{fig:tension}) nearly coincide with the model's prediction for deglycosylated aggrecan. As noted above, in cartilage the complex is often stabilized by link protein, which reduces the aggrecan spacing to $d \approx$10 nm. As $f_{int}$ depends sensitively on $d$, link protein's contribution will likely lead to a much higher tension in healthy physiological complexes: $\sim$2 pN, instead of $\sim$0.5 pN as reported here.

\subsection*{Aggrecan polydispersity}

The above model makes predictions for extreme cases: monodisperse populations of either undamaged aggrecan or completely deglycosylated core protein. It is clear that a monodisperse population of aggrecan with intermediate glycosylation (and thus $b_a$ between 1.6 nm and 50 nm) would generate a tension intermediate between these cases. In practice however, there will usually be a mixture of aggrecan at varying levels of glycosylation, ranging from the bare core protein to a dense bottlebrush. In the following, we consider a two-species mixture of undamaged and deglycosylated aggrecan. In neglecting intermediate levels of glycosylation, this scenario is still a significant simplification. Nonetheless, it illustrates how polydispersity in the aggrecan population can lead to variability in the extension of aggrecan-decorated HA chains.

We consider a simple model where the HA chain is represented by a 1D array of binding sites, spaced by $d = 26$ nm. Each site can be occupied either by an undamaged aggrecan (A) or a deglycosylated core protein (B). Each bound side chain interacts only with its nearest neighbors, stretching the intervening HA segment. If stretched by A's, the segment will experience $f_{int}^{AA} =$ 0.51 pN, as calculated above, while a segment stretched by two B's will experience $f_{int}^{BB} =$ 0.15 pN. We assume that the tension in a segment with one A neighbor and one B neighbor, $f_{int}^{AB}$, is closer to the completely deglycosylated case due to the lack of charged CS on the core protein, as well as its presumed ideal (i.e. self-intersecting) behavior. Thus we take $f_{int}^{AB} =$ 0.15 pN as well. We define $\Phi$ as the fraction of undamaged A aggrecan in the bulk solution, and populate each binding site stochastically with probability $P$(A)$ = \Phi$. We use the asymptotic form of the MS-WLC expression \cite{Marko1995} to predict the relative length of a segment as a function of the total force it experiences (external force + internal tension):
\begin{equation}
\frac{L^{seg}}{L_c^{seg}}=1-\sqrt{\frac{k_BT}{4(f+f_{int})l_p}}.
\label{eq:WLC_2}
\end{equation}
We count the number of each type of segment (A-A, A-B, and B-B) to approximate force-extension curves for HA decorated by a heterogeneous aggrecan population. 
 
In our model, variability can arise due to two factors. The quality of aggrecan may differ between samples; thus we make $\Phi$ a random variable. We will set the mean $\bar{\Phi}  =$ 0.8, and assume $\Phi$ is normally distributed with a standard deviation of 10\%. Additionally, even without noise in $\Phi$, the randomness inherent in the binding process leads to fluctuations in the \textit{actual} fraction of A aggrecans bound, as well as their arrangement on the chain. For short HA tethers (e.g. $\sim$600 nm, or $\sim$20 binding sites), these fluctuations from the average are significant. Our experimental setup invariably leads to polydispersity in tether length: the tethers included in Fig. \ref{fig:tension} range in length from $\sim$600 nm to $\sim$5000 nm ($\sim$200 binding sites). Thus, in our model, we select a contour length at random from the set of 10 experimentally measured tethers. 

\begin{figure}[h!]
\centering
\includegraphics[width = 3.25in]{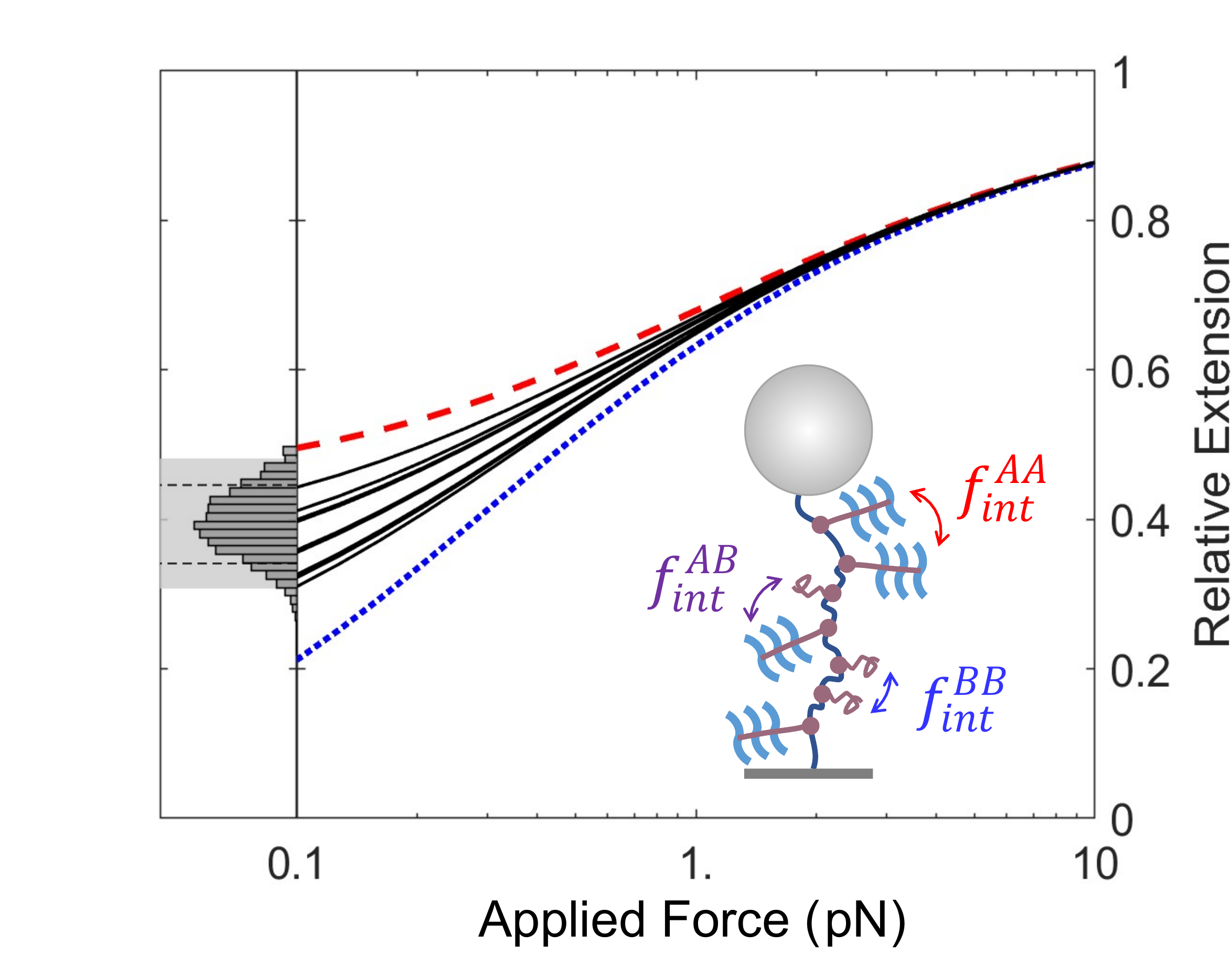}
\caption{Predicted force-extension curves for HA in a mixture of undamaged and deglycosylated aggrecan. Right: Red dashed (blue dotted) curve shows the predicted force-extension behavior for $\Phi =$ 1 (0). Black curves show predicted force-extension for 10 tethers, each populated randomly with whole aggrecan and deglycosylated core protein according to probability $\Phi$, with $\Phi$ drawn from a distribution of mean $\bar{\Phi} =$ 0.8. For each tether, $L_c$ is selected randomly as described in the text. Left: histogram shows the length at 0.1 pN for 100000 simulated tethers. The shaded gray region includes 95\% of the data. The dashed lines indicate how this region would narrow if there were no noise in $\Phi$. Inset: schematic illustrating the 1D binding model and the heterogenous tensions induced in segments throughout the chain.}
\label{fig:1Dmodel}
\end{figure}

Our model results qualitatively show that sample-to-sample variation in aggrecan quality, along with random fluctuations in the actual bound population and arrangement on short tethers, can lead to the kind of noisy data we observe experimentally. We simulate $10^5$ tethers, each with random $\Phi$ and $L_c$ as described above. Force-extension curves for the first 10 trials are plotted on the right side of Fig. \ref{fig:1Dmodel}. The histogram on the left of Fig. \ref{fig:1Dmodel} shows the low-force (0.1 pN) extension of all $10^5$ tethers. With no sample-to-sample variation ($\Phi$ fixed to 0.8), this distribution narrows significantly, suggesting that sample quality variation is needed to explain the data. Further, our experiments include two pairs of curves from tethers that were measured simultaneously and thus experienced the same aggrecan sample (same $\Phi$). These pairs of tethers show nearly identical mechanical responses (Fig. S5), suggesting that sample-to-sample differences in aggrecan quality are the major cause of the variability in the force-extension curves. 

\subsection*{HA structure on short length-scales}

Prior work has shown that some HA-binding proteins, such as the proteoglycan versican G1, bind cooperatively and may induce higher-order helical structure in HA \cite{Seyfried2006}. Such a structure would affect the HA chain on relatively short length-scales, and would be measured as a decrease in extension under relatively high tension. It is known that the aggrecan G1 domain binds to 5 disaccharide units \cite{Hardingham1973}, approximately equal to the intrinsic HA persistence length. Bends on this length-scale would likely manifest as a decrease in the fit persistence length \cite{Kulic2005}.

We see no evidence that aggrecan binding alters the short length-scale structure of HA by inducing bends or helical superstructure. At sufficiently high force (significantly greater than $\sim$1 pN, probing structural features smaller than $l_p$) we see no difference between force-extension curves with and without aggrecan (example: Fig. \ref{fig:data}C). Further, in agreement with Liu et. al. \cite{Liu2004again}, if we perform fits with persistence length as a free parameter, we do not find a statistically significant difference in fit HA persistence length with and without aggrecan when swelling is accounted for by an internal tension (bare HA: $l_p$ = 6.6 $\pm$ 0.4 nm; with aggrecan: $l_p$ = 6.1 $\pm$ 0.7 nm, N = 10, reported errors are standard error of the mean). We note that these estimates of the HA persistence length are consistent with prior magnetic tweezers experiments \cite{Berezney2017}, but are systematically larger than estimates from AFM \cite{Bano2018}; we attribute this to the general principle that high-tension stretching systematically biases measurements to lower persistence lengths \cite{Dobrynin2010}, e.g. by force disrupting local interactions and secondary structure in the chain.

To confirm this, we performed experiments using an HA-binding aggrecan fragment, G1-IGD-G2, which lacks the large CS-rich region that constitutes most of aggrecan's size and charge and is responsible for the large degree of swelling. Over the range of forces which probe short length-scale structure, there is no discernible difference in the response with and without G1-IGD-G2 (see example data in Fig. S7). Using SDS-PAGE, we qualitatively confirmed the fragment can bind to HA (Fig. S8). These results are in agreement with others who have found that unlike versican G1, aggrecan G1 does not bind cooperatively, does not pack tightly on HA, and does not cause the same superstructures \cite{Foulcer2012}. In vivo, aggrecan binds HA alongside link protein in a ternary complex \cite{Knudson2001}. It remains a possibility that link protein could induce higher-order structure in the HA at the center of these complexes, even if aggrecan binding alone does not, as it has been suggested that link protein induces cooperativity in aggrecan G1 binding \cite{Neame1993}.

\section*{Conclusion}

We establish low-force stretching as a useful technique to detect and study proteoglycan binding to HA. Side chain induced swelling affects HA's chain structure on long length-scales: 10s to 100s of nm. Structure on these scales is sensitive to sub-piconewton forces, accessible using magnetic tweezers. Here, we have used this technique to quantify the expansion undergone by a single HA chain at the center of an HA-aggrecan bottlebrush complex (see Fig. \ref{fig:data}), and found that inter-aggrecan repulsion generates an internal tension of about 0.5 pN. We have also examined the complexes under larger stretching forces ($\sim$1-10 pN) where all long length-scale structure has been pulled out and we are sensitive to small structures ($\sim$nm). At these high forces (short length-scales), we find no difference in HA chain structure with and without aggrecan, indicating that while aggrecan causes significant swelling of the random coil structure of HA, its binding does not cause any local structure changes. While all experiments show swelling, we observe significant variability in the force-response of the aggrecan-decorated HA chains, which is likely caused by polydispersity of the aggrecan. We use a theory of bottlebrush polymers with previously measured physical parameters for aggrecan, and find that our data are bounded by the expected limits for fully glycosylated and deglycosylated aggrecan. In doing so, we demonstrate the sensitivity of the magnetic tweezers technique to the molecular level degradation of proteoglycans, and connect the chemistry of aggrecan to the mechanics of larger-scale structures in cartilage.

Our experimentally validated model quantitatively predicts how bottlebrush architecture leads to the expanded nature of the HA-aggrecan complex, which is crucial in maintaining the structural integrity of cartilage. While swelling is expected from previous work \cite{Chang2016,Attili2013}, developing a firm physical basis for the system improves understanding of the basic processes, and offers quantitative methods that can later be extended to answer other biological questions. Aggrecan glycosylation undergoes slight changes with age, but the more dramatic alterations to aggrecan structure occur in pathologic conditions, where aggrecanase enzymes cleave the core protein, at times removing some or all of the CS-attachment domain \cite{Ng2003,Dudhia2005}. The model presented here makes predictions for how the complex structure will change in the case of reduced core protein size; future work will explore aggrecan proteolysis and evaluate these predictions. Further, the internal tension/brush model will likely be useful to describe complexes of HA with other proteoglycans (e.g. versican and brevican), which have a similar bottlebrush geometry, but drastically different biological functions. In these various complexes, changes in the bottlebrush structure will set the tension; this in turn will affect not only the extracellular matrix mechanics, but also the force signals transmitted from the exterior to the interior of cells, by means of the mechanosensitive HA receptor CD44 \cite{Kim2014}.

\section*{Author Contributions}

O.A.S., J.P.B, and S.I.-G. designed the research. S.I.-G. and J.P.B. designed and performed experiments, and analyzed data. S.I.-G., J.P.B., and O.A.S. wrote the article. 

\section*{Supporting Material}
Data supporting experimental results and choice of model parameters, Figures S1-S8. 

\section*{Acknowledgments}
This work was supported by the National Science Foundation under grants no. 1611497 and no. 2005189. S.I.-G. acknowledges support from a Connie Frank Fellowship.
\\
\\
This work is licensed under the Creative Commons Attribution-NonCommercial-NoDerivatives 4.0 International License (CC BY-NC-ND). To view a copy of this license, visit 
\\
https://creativecommons.org/licenses/by-nc-nd/4.0/.



\section*{Supplementary material}

Supplemental Information for: "Single-molecule stretching shows glycosylation sets tension in the hyaluronan-aggrecan bottlebrush." S. N. Innes-Gold, J. P. Berezney, and O. A. Saleh.

## I. Dependence of HA swelling on aggrecan concentration

When small amounts of aggrecan are added to solution, increases in the aggrecan concentration lead to increases in HA extension, presumably due to an increase in the number of bound aggrecans. However, with further increases in aggrecan concentration, little additional swelling occurs. We find that beyond about ~0.5 to 1 mg/mL aggrecan, increasing concentration cannot force more aggrecan onto the HA chain, and extension changes level off (Fig. S1). We perform our experiments with a solution concentration of 2 mg/mL, well into the concentration-insensitive regime.

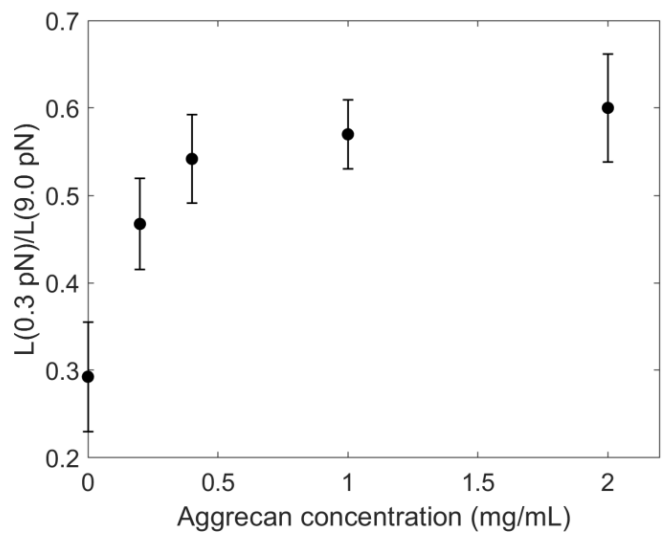

Figure S1. Extension of HA tethers at an applied force of 0.3 pN normalized by extension of the same tether at 9 pN, plotted versus solution concentration of aggrecan. Measurements are an average over 7 tethers. Error bars show standard deviation.

## II. Verification of presence and concentration of aggrecan core protein.

Aggrecan from Bovine Articular Cartilage, MW > 2.5 MDa, purchased from Sigma-Aldrich, was digested with chondroitinase ABC (chABC; Recombinant *P.vulgaris* Chondroitinase ABC with N-terminal Met and 6-His tag, purchased from R&D Systems). The digestion protocol and subsequent removal of chABC are described in the main text. After digestion, the presence of deglycosylated core protein at approximately the expected concentration was confirmed using SDS-PAGE. The right-most columns of Fig. S2 show a band at approximately 300 kDa, as expected for aggrecan core protein. The chondroitin sulfate side chains likely make whole aggrecan too large to enter into the gel, and with undigested aggrecan, the ~300 kDa band was not observed (data not shown).

17

Columns 1-4 in the gel in Fig. S2 represent increasing amounts of core protein. Assuming all aggrecan was completely digested, and that no core protein was lost during the digestion and enzyme removal, and assuming the core protein constitutes about 10% of the whole aggrecan molecular weight, we expect to have 10, 20, 30, and 50 ng of core protein respectively in columns 1-4. Biosafe Coomassie Dye has a lower detection limit of 8-29 ng per band. The aggrecan core protein band (~300 kDa) is clearly visible in columns 2, 3, and 4. We therefore conclude that the core protein is present at approximately the expected concentration.

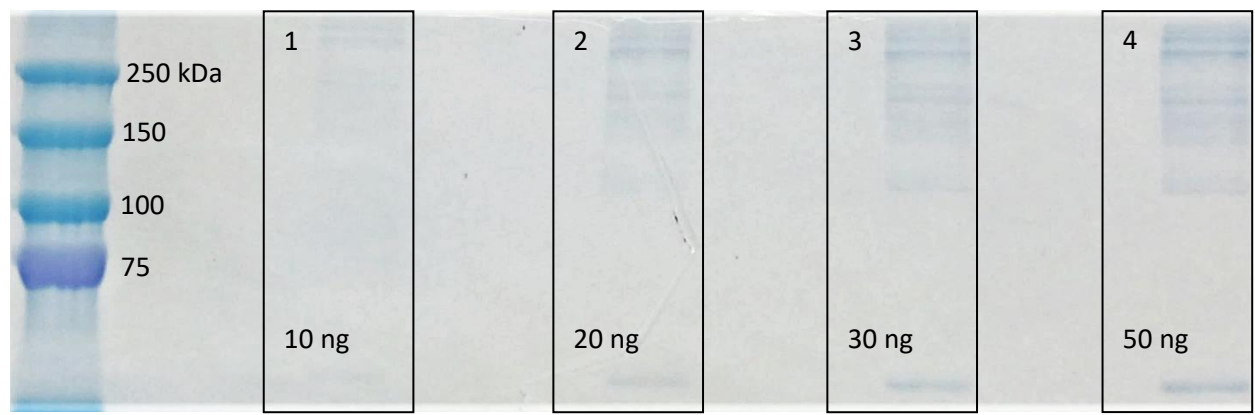

Figure S2. PAGE gel on digested aggrecan. Columns 1-4 have increasing amounts of core protein.

We performed dynamic light scattering (DLS) on aggrecan before and after digestion with chABC. DLS studies used a Zetasizer Nano ZS system from Malvern Panalytical. DLS measurements of aggrecan in denaturant-free buffer suffered from poor measurement quality, likely due to aggregation. To resolve this, DLS measurements on both whole and digested aggrecan were performed in 4M Guanidinium HCl to prevent aggregation. Based on the length of the aggrecan core protein (as described in main text) we expect whole aggrecan to be on the order of 100 nm in size, whereas we expect core protein alone, no longer extended by chondroitin sulfate (CS) side chains, to have a hydrodynamic size on the order of 10 nm. Figure S3 shows DLS results before and after digestion, confirming that this is the case.

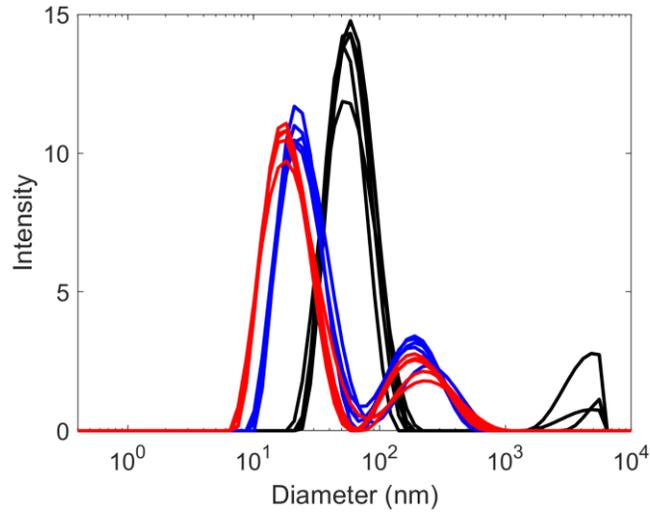

Figure S3. DLS measurements on aggrecan, before and after digestion with chABC. Black: Undigested aggrecan, peak at 59 nm. Blue: Digested for 2.5 hours, peak at 21 nm. Red: Digested for 27 hours, peak at 18 nm. All samples were measured in 4 M guanidinium chloride denaturant.

### III. Free chondroitin sulfate has no effect on hyaluronan elasticity.

The enzyme chABC has activity on HA as well as on CS. In addition to the chABC removal protocol described in the main text, free CS (chondroitin sulfate sodium salt from shark cartilage, MW 50-59 kDa, purity >95%, purchased from Sigma-Aldrich) was included in the buffer for HA pulling experiments to protect the tethers from any remaining enzyme. We performed control force-extension measurements to demonstrate that including 1 mg/mL free CS has no effect on the HA tethers. Representative data are shown in Fig. S4.

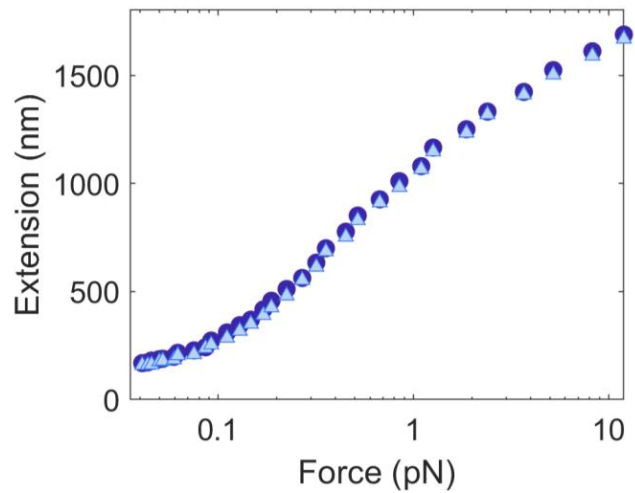

Figure S4. Example force-extension curves on a single HA tether in CS-free buffer (dark blue circles) and buffer containing 1 mg/mL CS (light blue triangles). In both cases, the chain displays an identical mechanical response.

## IV. Tethers measured simultaneously exhibit similar behavior.

Despite the large scatter observed in the force-extension response of HA tethers measured in 2 mg/mL aggrecan (Fig. 3, main text), we observe that tethers measured simultaneously exhibit very similar behavior (Fig. S5). Although the number of samples is small (two pairs of curves), this suggests that much of the noise is due to sample quality differences between experiments, rather than due to the randomness inherent in binding for a polydisperse aggrecan population.

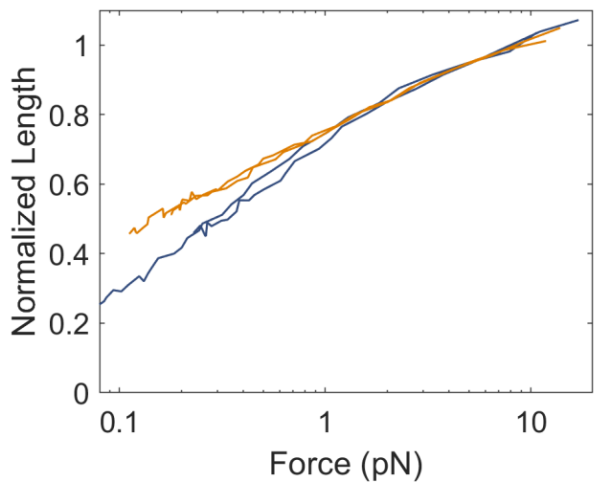

Figure S5. Two pairs of force-extension curves on HA tethers. Orange and blue denote separate experiments, performed on different days with different samples of aggrecan (both 2 mg/mL).

## V. Theoretical predictions using different assumptions.

In applying the polymer physics theory developed by Panyukov *et. al.*[1] (see main text for details) to predict changes in the HA-aggrecan bottlebrush upon deglycosylation, we make several assumptions regarding the solution structure of the aggrecan core protein.

In the text, we assumed a minimal Kuhn length of 1.6 nm for deglycosylated aggrecan, expected for a fully unfolded polypeptide.[2] Others have shown that some regions of aggrecan are well-described by a slightly longer Kuhn length of around 3.5 nm.[3] With this longer Kuhn length, the internal tension is predicted to increase slightly, to 0.20 pN (rather than 0.15 pN). The resulting force-extension curve is shown in green in Fig. S6. We also assumed that the deglycosylated core protein behaves ideally, following studies of other predominantly disordered proteins.[4] If we instead assume swollen chain behavior, we find an internal tension of 0.21 pN (force-extension prediction shown in yellow in Fig. S6). Ultimately, for physically reasonable choices of parameters, the model results do not change drastically, and our conclusions are unaffected.

20

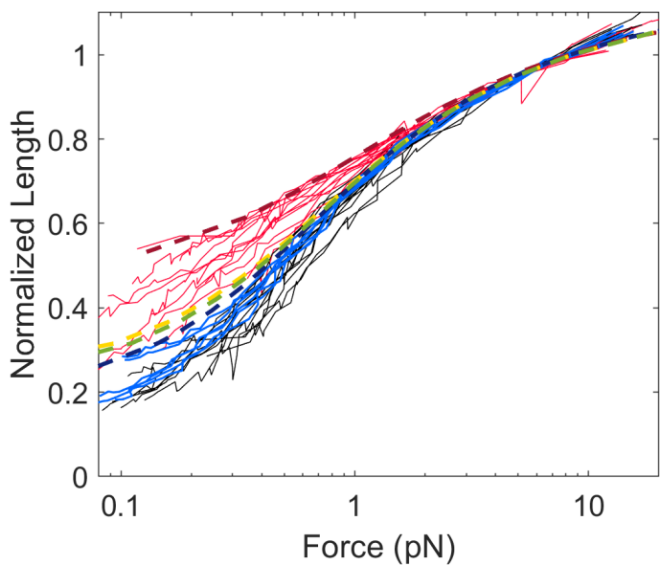

Figure S6. Effect of different assumptions on the bottlebrush physics model. Overlaid on Fig. 3, we show the force extension curves predicted for deglycosylated aggrecan with a Kuhn length of 3.5 nm (green dashed curve), and for deglycosylated aggrecan that behaves as a swollen chain (yellow dashed curve). As discussed in the main text, the blue dashed curve assumes a Kuhn length of 1.6 nm and ideal scaling.

## VI. Aggrecan binding does not alter short length-scale HA chain structure.

We perform force-extension measurements on HA in the presence of a recombinant HA-binding aggrecan fragment (Recombinant aggrecan G1-IGD-G2 with 10-His tag, purchased from BioLegend) that does not include the large, CS-rich domain of aggrecan. Example HA force-extension curves, with and without the aggrecan fragment, are shown in Figure S7. As discussed in the main text, binding-domain induced structure changes (e.g. bends or kinks) would likely affect the chain on length scales below ~5 nm, and would manifest as a change in elasticity at forces larger than ~1 pN. We see no evidence of changes to HA structure in this regime. We observe a slight shortening of the chain at low forces (below ~0.5 pN); this is likely due to the large, positively-charged 10-His tagged on the HA-binding fragment, which may shorten the negatively charged HA chain by screening intra-chain electrostatic repulsion.

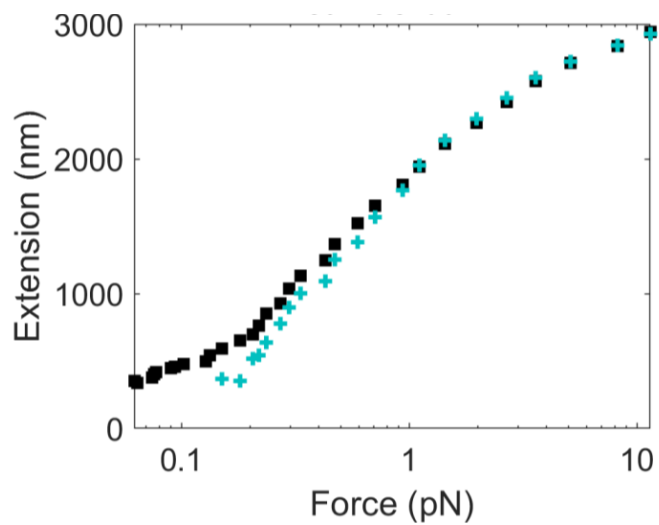

Figure S7. Example force-extension curves on a single HA tether with (green crosses) and without (black squares) 0.8 µM G1-IGD-G2 aggrecan fragment.

We performed the following experiment to qualitatively verify that the recombinant G1-IGD-G2 aggrecan fragment binds to HA in the experimental buffer conditions (100 mM NaCl, 10 mM MOPS, pH 7). We placed samples of the aggrecan fragment into two 100 kDa spin columns, A and B. Sample A contained only the protein, while to sample B we added high-MW HA (Sigma-Aldrich, 1.25-1.5 MDa) with binding sites in at least 100-fold excess over the added protein. We centrifuged both samples. For sample A, with no HA, it is expected that much of the 74.3 kDa protein will pass through the 100 kDa filter into the filtrate. If HA (which is too large to pass through the filter) retains bound protein, we would expect filtrate B to contain much less protein than filtrate A. The filtrates were analyzed with SDS-PAGE. In the gel (Fig. S8), a strong protein band is clearly visible for the sample A filtrate. For the sample B filtrate, a very faint band is just barely discernable, suggesting that the HA caused most of the protein to be retained in the concentrate.

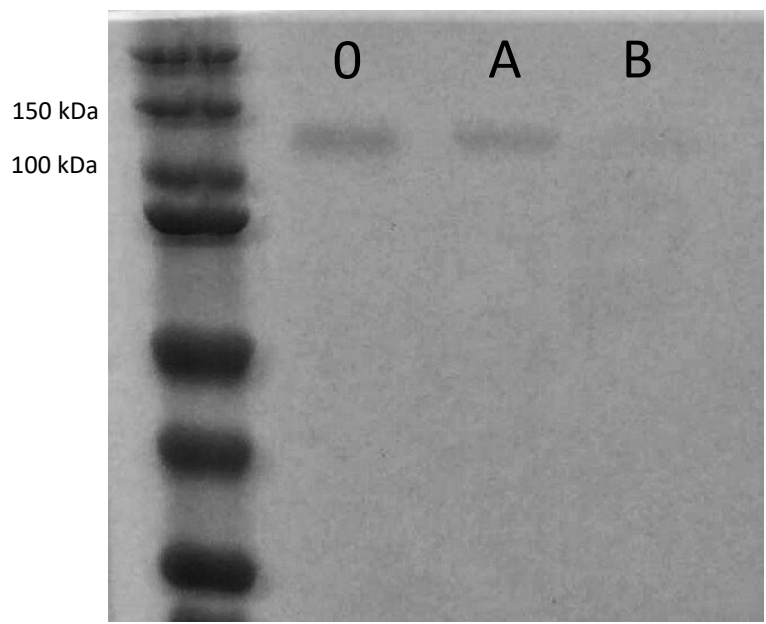

Figure S8. SDS-PAGE on recombinant G1-IGD-G2 protein with 10-His tag. Lane 0 shows the protein with no filtration steps. Despite its size of 74.3 kDa, the protein migrates at around 100 kDa in SDS-PAGE. Lane A demonstrates that much of the protein passed through the 100 kDa spin filter into the filtrate, while lane B shows that if high-MW HA is placed in the filter with the protein, very little protein passes through to the filtrate.



\end{document}